\documentclass[]{interact}

\usepackage{epstopdf}
\usepackage{subfigure}
\usepackage{upgreek}
\usepackage{xcolor}
\usepackage{soul}

\usepackage[numbers,sort&compress,merge]{natbib}
\bibpunct[, ]{(}{)}{,}{n}{,}{,}

\soulregister\ref{7}
\soulregister\cite{7}
\soulregister\footnote{8}

\theoremstyle{plain}

\theoremstyle{definition}

\theoremstyle{remark}

\begin{document}

\articletype{ARTICLE DRAFT}

\title{Single-ion, transportable optical atomic clocks}

\author{
\name{Marion Delehaye and Cl\'ement Lacro\^ute\thanks{CONTACT Cl\'ement Lacro\^ute Email: clement.lacroute@femto-st.fr} }
\affil{FEMTO-ST institute, univ. Bourgogne Franche-Comt\'e, CNRS, ENSMM, Time and frequency dept., 26 Rue de l'\'Epitaphe, 25030 Besan\c{c}on cedex, France.}
}

\maketitle

\begin{abstract}
For the past 15 years, tremendous progress within the fields of laser stabilization, optical frequency combs and atom cooling and trapping have allowed the realization of optical atomic clocks with unrivaled performances. These instruments can perform frequency comparisons with fractional uncertainties well below $10^{-17}$, finding applications in fundamental physics tests, relativistic geodesy, and time and frequency metrology. Even though most optical clocks are currently laboratory setups, several proposals for using these clocks for field measurements or within an optical clock network have been published, and most of time and frequency metrology institutes have started to develop transportable optical clocks. For the purpose of this special issue, we chose to focus on trapped-ion optical clocks. Even though their short-term fractional frequency stability is impaired by a lower signal-to-noise ratio, they offer a high potential for compactness: trapped ions demand low optical powers and simple loading schemes, and can be trapped in small vacuum chambers. We review recent advances on the clock key components, including ion trap and ultra-stable optical cavity, as well as existing projects and experiments which draw the picture of what future transportable, single-ion optical clocks may resemble. 
\end{abstract}



\section*{Introduction}

Outstanding performances have been obtained by optical atomic clocks: single-ion clocks have achieved fractional frequency instabilities and accuracies as low as $3\times 10^{-15} \tau^{-1/2}$ for the former \cite{Chou2010} and $3\times 10^{-18}$ for the latter \cite{Huntemann_2016}, while optical lattice clocks have reached instabilities and inaccuracies as low as $6\times 10^{-17} \tau^{-1/2}$ \cite{Schioppo_2016} and $2 \times 10^{-18}$ \cite{Nicholson2015} respectively. Optical clocks are thus extremely powerful tools, to be used in various domains of precision science, and many proposals require leaving the well-controlled laboratory environment to bring optical clocks into the field. For this purpose, a number of laboratories worldwide are now building transportable optical clocks. The first operational setups have been reported recently, for both a strontium lattice clock \cite{Poli2014,Koller2017} and a single-ion calcium clock \cite{Cao2017}.

Optical lattice clocks benefit from their high number of atoms that leads to a reduced fractional instability, even though the constant reloading required induces an increased susceptibility to the Dick effect \cite{Quessada2003}. They also imply a complex loading scheme into the optical lattice trap, starting with a 2D magneto-optical trap (MOT) or a Zeeman slower, and including several intermediate MOTs and/or dipole traps. On the other hand, single-ion clocks involve a much simpler loading scheme and require lower optical powers, which can lead to more compact and less power-consuming setups.

A number of very detailed reviews on optical clocks were published in the past few years \cite{Margolis2010, Poli2013,Ludlow2015,Hong2017}. For this special issue in memory of Professor Dany Segal, we review emerging experiments and techniques useful for the realization of transportable, single-ion clocks.

We first mention several applications that could benefit from the use of transportable single-ion optical clocks. We then provide a short overview of single-ion clocks basics. Focus is then given to compact clock lasers and optical cavities, and to compact ion trap design considerations. We finally review the current existing transportable clock projects before targeting specific developments that will help further reducing the clock footprint.

\section{Motivations}

\subsection{Internet of clocks}
Optical clocks are now much more stable and exhibit lower systematic uncertainty than the best cesium clocks that currently define the time unit.

Many laboratories and astronomical observatories would benefit from an access to those frequency references. 
For instance, VLBI (Very Long Baseline Interferometry) relies on the simultaneous observation of various radio signals by distant antennas. 
These signals are down-sampled using a local oscillator, the stability of which is of paramount importance in order to perform low-noise correlations and access high angular resolution.
The local oscillators are usually synchronized to active hydrogen masers; a test using a better reference was conducted using a cryogenic sapphire oscillator \cite{Nand2011}. However, the need for even better local oscillator stability has emerged recently \cite{Clivati2015,Przemyslaw2016}. The first synchronization to a remote atomic clock located in a National Metrology Institute has been reported \cite{Clivati2017}, and is paving the way for synchronization to an optical clock.  This would give access to investigation of compact radio sources or molecular emission from the interstellar medium.

A generalized access to optical frequency references and clock comparisons is also a prerequisite to the redefinition of the SI second from optical clocks \cite{Gill_2016} that will be rediscussed in the early 2020s and would impact other SI units \cite{Mills_2006}. GPS and satellite links, with a relative frequency stability already remarkable, on the order of $10^{-15}$ \cite{Leute2016} and $3\times 10^{-16}$ \cite{Exertier2010,Fujieda2014} per day respectively, cannot keep up with the performances of the best optical clocks. 
Optical fibers are now widely used for optical and RF frequency dissemination over long distances \cite{Musha2008, Predehl2012, Droste2013, Sliwczynski2013} and have reached fractional frequency stabilities of $4\times 10^{-16}$ at 1 s, averaging down to $10^{-19}$ in 2000~s \cite{Chiodo2015}, but their installation is costly and time-consuming, and a regular array at a continental scale is hard to contemplate in the mid-term. Transportable optical clocks would provide local and/or temporary links to the emerging optical clock network \cite{Riehle2017}.

\subsection{Geodesy}
Transportable optical clocks are at the heart of a new approach to geodesy called chronometric leveling \cite{Vermeer1983,Bjerhammar1985,PacomeDelva2013,Flury2016}. It is based on the fact that the relative frequency shift between two identical clocks located at different gravity potentials is given by $\frac{\Delta\nu}{\nu_0}=\frac{\Delta U}{c^2}$, where $c$ is the speed of light and $\Delta U$ the gravity potential difference. For two optical clocks near the Earth surface, this relative shift is about $10^{-16}$ per meter height difference. With uncertainties on systematic shifts below $10^{-18}$, this implies that the geoid (\emph{ie} the equipotential of the geopotential closest to mean sea level) can be measured with a centimetric precision at a local scale. 
Height and geopotential differences between several national metrology laboratories have been measured accurately~\cite{Predehl2012, Lisdat2016, Takano2016, Nemitz_2016}, but mobile atomic clocks are aimed to play a key role in the determination of the geopotential with increased spatial resolution~\cite{Vermeer1983, Lion2017}. The first on-field chronometric leveling measurement campaign using a transportable lattice clock has been reported recently \cite{Grotti2017} and its results are in excellent agreement with independent geodetic measurements.

\subsection{Fundamental physics}
Being such precise tools, optical clocks are the key to challenge and probe the faintest phenomena in fundamental physics (see for instance \cite{Safronova2017} for a review). 
Some theoretical predictions infer for instance that the yet-elusive nature of dark matter could be probed with a continental network of atomic clocks \cite{Derevianko_2014,Derevianko_2016}.
Many efforts are also dedicated to the search of a violation of Einstein Equivalence Principle \cite{Damour2002}. Atomic clocks mainly address the issue of a potential breakdown of Local Position Invariance \cite{Uzan2003}, through tests of both the invariance of fundamental constants and the universality of gravitational redshift. The former focuses on the search of variations of the fine structure constant $\alpha$ \cite{Godun2014,Dzuba_2016} or the $m_p/m_e$ ratio \cite{Huntemann2014} while the latter requires the comparison of two clocks undergoing different gravitational potentials and will greatly benefits from the perspectives of sending atomic clocks in space, with notably the ACES/PHARAO mission \cite{Laurent2015} that is scheduled to be launched within the year 2018. 


\subsection{Towards space applications}
The above applications rely at least partly on the perspective of sending optical clocks in space. After various proposals (STE-QUEST\footnote{Space-Time Explorer and Quantum Equivalence Principle Space Test}, SAGAS\footnote{Search for Anomalous Gravitation with Atomic Sensors}, EGE\footnote{Einstein Gravity Explorer}), the SOC2 (Space Optical Clocks) project has been launched by the European Space Agency (ESA), with current funding from the European Union Seventh Framework Program. 
Even though the SOC2 project is not based on ion clocks but rather on Sr and Yb optical lattice clocks \cite{soc2,Bongs2015,Origlia2016,Pizzocaro2017}, it emphasizes the growing need for optical clocks in space.

The presence of master clocks in space would improve deep-space navigation and allow direct remote comparisons between on-ground high-accuracy optical clocks. A constellation of optical clocks in space would also enhance the GNSS system by reducing its dependence on regular updates from ground stations. 

In addition to constraints such as a high resistance to vibrations and radiations, space systems imply a reduction of size, weight and power consumption. In this respect, the compact and transportable optical clocks that will be discussed in this paper are thus a precursor for space optical clocks.
\\


\subsection{State of the art}
Following early proposals by H. Dehmelt \cite{Dehmelt1975,Dehmelt1989,Yu1992}, single-ion optical clock experiments started in the 1990s \cite{Bell1991}, and since the appearance of optical frequency combs in the early 2000s \cite{Cundiff2003, Hansch2006}, tremendous progress has been achieved in the field (see for instance \cite{Margolis2010,Poli2013,Ludlow2015,Hong2017} for reviews). Laboratory single-ion optical clocks have been developed in nearly all the main national metrology institutes and time and frequency laboratories, and at least five different species have been extensively tested and characterized. Table~\ref{table:soa} shows most of these species, and lists their clock transition wavelength, linewidth, best-accuracy to date, and laboratories.

\begin{table}[b!]
\tbl{State of the art of laboratory single-ion optical clocks. Laboratory acronyms: National Institute of Standards and Technology (NIST, USA), National Physical Laboratory (NPL, UK), National Research Council (NRC, Canada), Physikalisch-Technische Bundesanstalt (PTB, Germany), Wuhan Institute of Physics and Mathematics (WIPM, PRC).}
{\begin{tabular}{lccccc} \toprule
 Ion			 										& Clock transition wavelength 	& Clock transition linewidth 	& Best measured frac. syst. unc. 							& Laboratories	\\ \midrule
 $^{27}$Al$^+$								& 267 nm 												&	8 mHz												& $8.6{\times}10^{-18}$ \cite{Chou2010} 			& NIST, PTB 		\\
 $^{40}$Ca$^+$ 								& 729 nm 												& 140 mHz 										& $5.7{\times}10^{-17}$ \cite{Huang2016}			& WIPM 					\\
 $^{199}$Hg$^+$							 	& 282 nm	 											& 1.8 Hz 											& $1.9{\times}10^{-17}$ \cite{Rosenband2008}	& NIST  				\\
 $^{88}$Sr$^+$								& 674 nm 												& 400 mHz 										& $1.2{\times}10^{-17}$ \cite{Dube2014}				& NPL, NRC 		\\
 $^{171}$Yb$^+$, quadrupole		& 436 nm 												& 3.1 Hz 											& $1.1{\times}10^{-16}$ \cite{Tamm2014}				& NPL, PTB  		\\
 $^{171}$Yb$^+$, octupole			& 467 nm 												& $\approx$ nHz 							& $3.2{\times}10^{-18}$ \cite{Huntemann_2016}	& NPL, PTB 			\\ \bottomrule
\end{tabular}}
\label{table:soa}
\end{table}

The record systematic uncertainty has been obtained by the Physikalisch-Technische Bundesanstalt (PTB) with the electric octupole transition at 467 nm in Yb$^+$, with a fractional value of $3{\times}10^{-18}$ \cite{Huntemann_2016}. Al$^+$, Hg$^+$ and Sr$^+$ have also demonstrated fractional systematic uncertainties around $10^{-17}$ or below \cite{Chou2010, Rosenband2008, Dube2014}, and several experiments now target the 10$^{-19}$ range using In$^+$ \cite{Pyka2013}, Th$^{3+}$ \cite{Peik2003} or Lu$^+$ \cite{Arnold2015}.

\section{Single-ion optical clocks: working principle}

\subsection{Single-ion optical clockwork}

Atomic clocks all rely on the same scheme: the output frequency of a local oscillator is locked to the frequency of an energy transition of an atomic reference. $^{133}$Cs hyperfine transition at $9.192 631 770 \ \rm{GHz}$ defines the SI second. Many other atoms and ions have been used in atomic clocks, chosen on criteria ranging from the reference frequency value to experimental ease-of-use. 
Thanks to their tremendous performances, their conceptual simplicity and the accumulated knowledge gathered in the field of ion-trapping during the past fifty years, single-ion optical clocks play a major role in the field of time and frequency metrology.

\begin{figure}%
\includegraphics[width=\columnwidth]{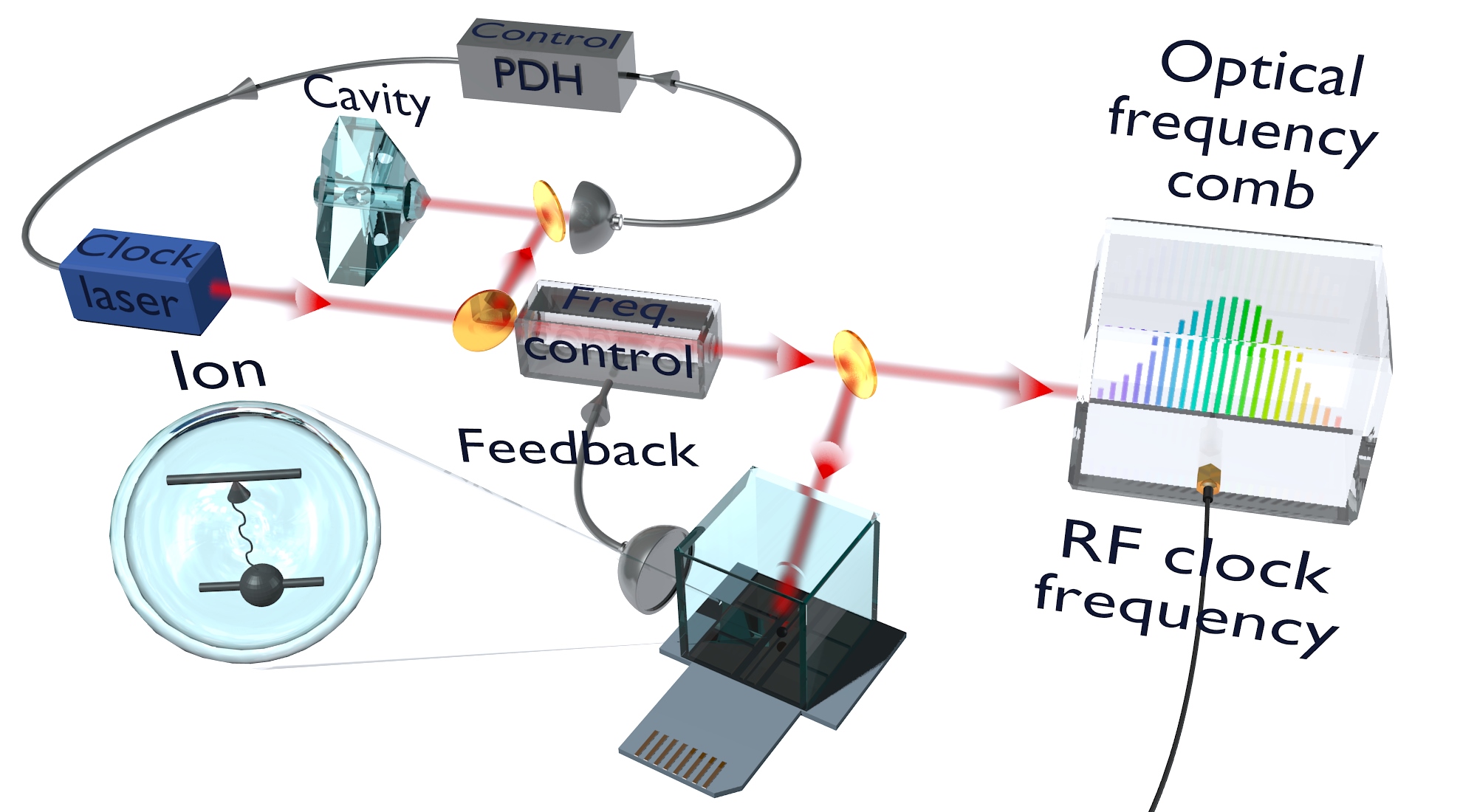}%
\caption{Schematic single-ion clockwork (see main text for details). PDH: Pound-Drever-Hall.}%
\label{fig:clockwork}%
\end{figure}

The schematic single-ion clockwork is illustrated Fig. \ref{fig:clockwork}. The clock laser frequency is compared to an optical transition frequency in a single, laser-cooled trapped ion and corrected using an optical frequency corrector such as an acousto-optical modulator (AOM). The output of this corrector is the optical clock output signal, which is locked to an ultra-stable Fabry-P\'erot (FP) cavity at short times ($\approx 1 \upmu$s - 10 s, see section \ref{sec:clocklaser}) and to the ion at longer times. It constitutes an optical frequency standard, which can be used and distributed directly for applications such as optical frequency comparisons, precision spectroscopy, relativistic geodesy, etc. Using an optical frequency comb, it can also be transferred to the radiofrequency (RF) domain, where it can form the basis of a timescale \cite{LeTargat2013, Grebing2015}. In that regard, the useful signal of an optical atomic \emph{clock} is actually the RF signal, which is why the optical frequency comb used to transfer the stabilized frequency from the optical to the RF domain should be taken into consideration in the overall setup volume (see section \ref{sec:comb}).


\subsection{Atomic transition}
The performance of an atomic clock is evaluated by its fractional frequency stability $\sigma_y$ and its accuracy, defined as the uncertainty on the average offset between the atomic transition frequency reference (at rest at a temperature of 0 K) and the clock output frequency. The ionic reference will therefore be chosen in order to provide a good fractional frequency stability, which requires a narrow transition linewidth, and a low sensitivity to external effects that can perturb the clock frequency, such as Zeeman or light shifts.

The ultimate fractional frequency stability of a single-ion clock is expressed as (see \cite{Champenois2004} for instance):
\begin{equation}
\sigma_y (\tau) = \frac{\delta \nu}{\nu_0} \frac{1}{S\!N\!R}\sqrt{\frac{T_c}{\tau}},
\label{eq:stab}
\end{equation}
where $\nu_0$ is the reference frequency, $\delta \nu$ is the observed linewidth, $S\!N\!R$ is the signal-to-noise ratio, $T_c$ is the clock cycle time, and $\tau$ is the integration time. For a single-ion clock limited by the quantum projection noise \cite{Itano1993}, the best achievable signal-to-noise ratio and linewidth depend on the excitation scheme, with $S\!N\!R=\sqrt{\frac{1-p}{p}}\approx 1$, where $p$ is the probability to excite the atom to the clock upper state \cite{Itano1993, Champenois2004}. $\delta \nu$ is ultimately limited by the transition natural linewidth. In single-ion optical clocks, narrow linewidths are found using forbidden transitions, such as quadrupole transitions in one-valence electron systems (Ca$^+$, Hg$^+$, Sr$^+$, Yb$^+$), octupole-electric transitions (Yb$^+$), or transitions between states with no electric quadrupole moment (\emph{eg} ${}^1\mathrm{S}_0 \rightarrow {}^3\mathrm{P}_0$ in Al$^+$ or In$^+$).

Several systematic effects shift and/or broaden the atomic transition, possibly impacting both the clock fractional frequency stability and accuracy:

\begin{itemize} 
	\item the \emph{Doppler shifts}, directly caused by ion motion. The first-order Doppler shift is canceled by operating in the Lamb-Dicke regime \cite{Dicke1953} by laser-cooling the ion inside a tightly confining trap. The second-order Doppler shift or time-dilation shift is a relativistic effect expressed as $\delta\nu_{td}/\nu=v^2/(2c^2)$, where $v$ is the ion velocity and $c$ is the speed of light in vacuum. The contribution of the atom thermal motion to this effect is strongly reduced by laser-cooling \cite{Chen2017}. The so-called micromotion driven by the RF trapping field is the main contribution to this shift, and can only be minimized by placing the ion as close to the RF null as possible \cite{Berkeland1998,Keller2016}.
	\item the \emph{electric quadrupole shift}, which induces a sensitivity to electric field gradients for the electronic states with non-zero electric quadrupole moment. Among the species under study, only Al$^+$ and In$^+$ have no electric quadrupole shift. When the quadrupole moment is non-zero, the clock transition will be shifted by any electric gradient, some of which cannot be measured to a high degree of precision (\emph{e.g.} patch potentials, which slowly vary in time). Two techniques can be used to effectively reduce the quadrupole shift fractional contribution below $10^{-17}$: one can average clock measurements along three orthogonal magnetic field orientations \cite{Itano2000}, or average measurements over several different $m_F$ transitions \cite{Dube2005}.
	%
	\item \emph{Stark shifts}, induced by the differential clock states polarizabilities. The ion is exposed to several electric fields, including the trapping potential (through micromotion), blackbody radiation (BBR) from the ion surrounding,  the clock laser, and in some cases a cooling laser (light shifts).
	\begin{itemize}
		\item The AC Stark shift induced by the ion micromotion can be minimized in the same manner as the second-order Doppler shift; for some species, the RF driving frequency can even be chosen such that the two shifts cancel out \cite{Itano2000, Dube2005}.
		\item The BBR shift depends on the differential polarizability of the clock states, and therefore varies significantly among the species. It will be lowest for Al$^+$ and In$^+$, which also have no electric quadrupole moment, placing them among the best candidates in terms of accuracy. The BBR shift contribution is nowadays reduced to $10^{-18}$ or lower for single-ion optical clocks \cite{Dolezal2015}. This was made possible by a detailed analysis of the ion surroundings temperature distribution, combining material analysis, temperature measurement and finite-elements-method (FEM) simulations \cite{Dolezal2015}.
		\item The light shifts are usually negligible, except in the case of the highly forbidden octupole transition in $^{171}$Yb$^+$ where high intensities are required to excite the ultra-narrow clock transition. For such cases, refined probing schemes can be used to cancel out the sensitivity to the probe laser intensity \cite{Zanon-Willette2016}. 
	\end{itemize} 
	\item the \emph{Zeeman shift}, which induces a sensitivity to external magnetic fields. This shift is usually canceled to first-order, either by using a $m_F=0 \rightarrow m_F=0$ transition (\emph{eg} for $^{199}$Hg$^+$ and $^{171}$Yb$^+$) or by averaging over hyperfine levels with opposite spins \cite{Bernard1998}.
\end{itemize}

Additionally to these metrological considerations, practical considerations cannot always be neglected. In the context of transportable clocks, the availability of laser sources to ionize, cool and interrogate the ion have an important impact on the setup size, weight and power consumption. Species involving transitions in the deep ultra-violet (UV) (below 300~nm), sympathetic cooling or quantum logic detection schemes, for instance, necessarily require a more complex optical setup. 
Some workarounds are being developed (see sections \ref{ssec:ptb} and \ref{ssec:nist}), aiming at the use of the most accurate candidates in future transportable clocks.

\section{Clock laser} \label{sec:clocklaser}
The clock laser used to interrogate the ion clock transition should ideally be narrower than its natural linewidth (ranging from below 1~mHz to a few Hz) and remain stable until
it can be locked to the atomic resonance, which is typically achieved with a few seconds timescale. 
This constraint is relaxed by the limited signal-to-noise ratio of single ion detection, which limits the fractional frequency stability of single-ion clocks to about $10^{-15} \tau^{-1/2}$ (using equation \ref{eq:stab} with realistic experimental parameters, see also \cite{Champenois2004}). In practice, the laser must be pre-stabilized with a fractional frequency stability well below $10^{-15}$ for integration times between $1- 30$ s. This is done using an ultra-stable, monolithic Fabry-P\'erot (FP) cavity, where the two mirrors defining the resonator are optically contacted to a spacer.

\subsection{Cavity design considerations}
The laser is stabilized to a Fabry-P\'erot resonator by locking its frequency to one of the cavity resonances using the Pound-Drever-Hall (PDH) technique \cite{Black2001}. Within the locking band, the cavity relative length stability $\sigma_{\delta L/L}$ is transferred to the laser relative frequency stability $\sigma_y$, as the FP resonances are all proportional to the free-spectral range $c/2L$, where $L$ is the FP length. Depending on the frequency noise of the laser which will be locked to the cavity, the PDH locking bandwidth might range from a few 10 kHz to a few MHz.

The ultimate performance of a FP resonator is limited by 
its length fluctuations at a finite temperature $T$, characterized by the thermal noise \cite{Numata2004, Kessler2012}.
The thermal noise can be lowered by working at low temperatures and/or by using materials with low mechanical losses, such as fused silica or single-crystal silicon. Crystalline highly-reflective coatings, which have a higher mechanical quality factor than traditional dielectric coatings, have also proved to reduce the thermal noise while still achieving high finesses \cite{Cole2013}.

In addition to the thermal noise, the cavity length is sensitive to external temperature and pressure fluctuations as well as accelerations. Ultra-stable FP resonators are therefore
thermally regulated, placed in ultra-high vacuum inside thermal shields and designed to be insensitive to accelerations. 
Moreover,  the cavity spacer on which the mirrors are contacted is usually made of a material with a vanishing coefficient of thermal expansion in order to reject temperature fluctuations to first order. While different materials can be used, including zerodur$^{\rm TM}$, ultra-low expansion glass (ULE), and single-crystal silicon \cite{Kessler2012a}, 
current transportable cavity designs are all based on ULE spacers.

\begin{figure}%
\centering
\includegraphics[width=0.4\columnwidth]{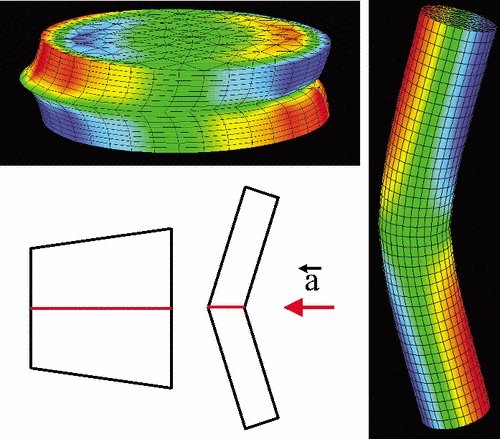}%
\caption{Example of an acceleration-insensitive design. Under transverse acceleration as shown by the red arrow, a thick vertical cylinder will experience a mirror tilt due to the Poisson effect, while a thin cylinder will bend and induce a tilt in the opposite direction. An aspect ratio can be found where these two effects will balance each other. From \cite{Millo2009}, copyright by the American Physical Society.}%
\label{fig:cavity_design}%
\end{figure}

The cavity sensitivity to accelerations is a result of its geometry and support forces. The general guiding principle is to maximize the cavity symmetries, and to use a design that makes the cavity length insensitive to accelerations and to the mounting forces \cite{Chen2006, Millo2009} (see Fig. \ref{fig:cavity_design} for an illustration). The two most common geometries are the vertical cylinder and the horizontal cylinder with square cutouts \cite{Millo2009}. In both designs, the support points and aspect ratios are chosen in such a way that the mirrors stay undeformed under accelerations. Several variations of these designs have been constructed to this date, now reaching fractional frequency stabilities below $10^{-16}$ \cite{Hafner2015, Matei2017}; however, in many of these, the cavity simply rests on its support and is held by its own weight, which is not possible in a transportable setup which can experience accelerations on the order of $g$ or higher.

\subsection{Transportable cavity designs}
\begin{figure}
\centering
\subfigure[Cubic cavity design, with 5 cm cavity length. The cavity is held at four of the eight corner cuts, resulting in a tetrahedral mounting with a null resulting force.  From \cite{Webster2011}.]{
\resizebox*{5cm}{!}{\includegraphics{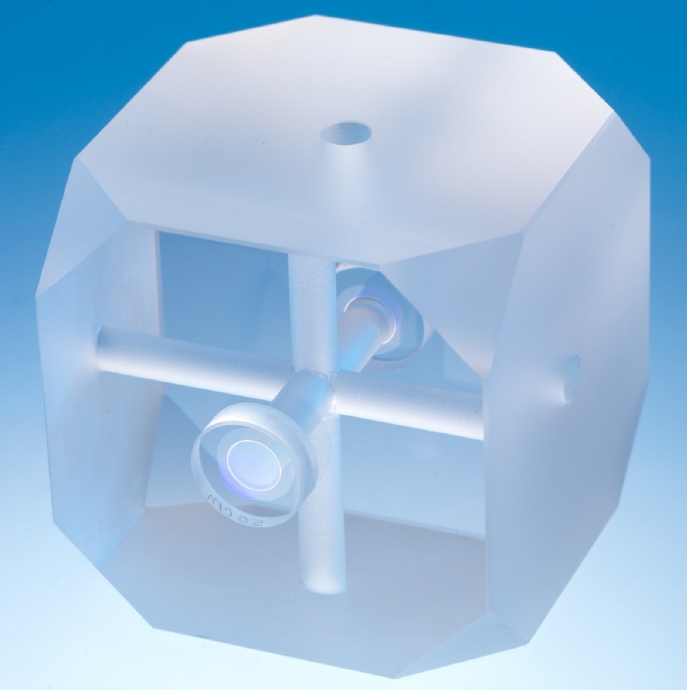}}}
\hspace{.3cm}
\subfigure[5 cm long spherical cavity in its vacuum chamber. The two-points mounting angle is chosen to reduce sensitivity to the support forces.]{
	\resizebox*{6.2cm}{!}{\includegraphics{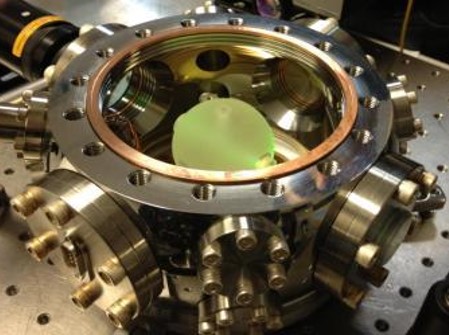}}}\hspace{5pt}
\subfigure[CAD rendering of a 2.5 cm long cylindrical cavity. The cavity is held on its faces using six teflon balls, placed in order to minimize the sensitivity to the support forces.  From \cite{Davila-Rodriguez2017}.]{
\resizebox*{5cm}{!}{\includegraphics{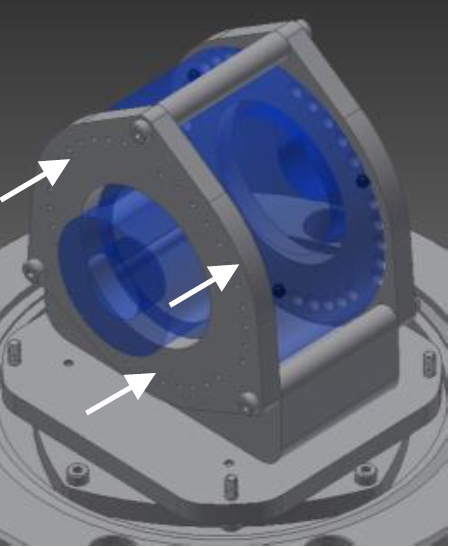}}}
\hspace{.3cm}
\subfigure[2.5 cm long double-tetrahedron ULE cavity at FEMTO-ST. The cavity is held rigidly in its mid-plane using three plane cutouts, inside a stainless steel holder.]{
	\resizebox*{6.3cm}{!}{\includegraphics{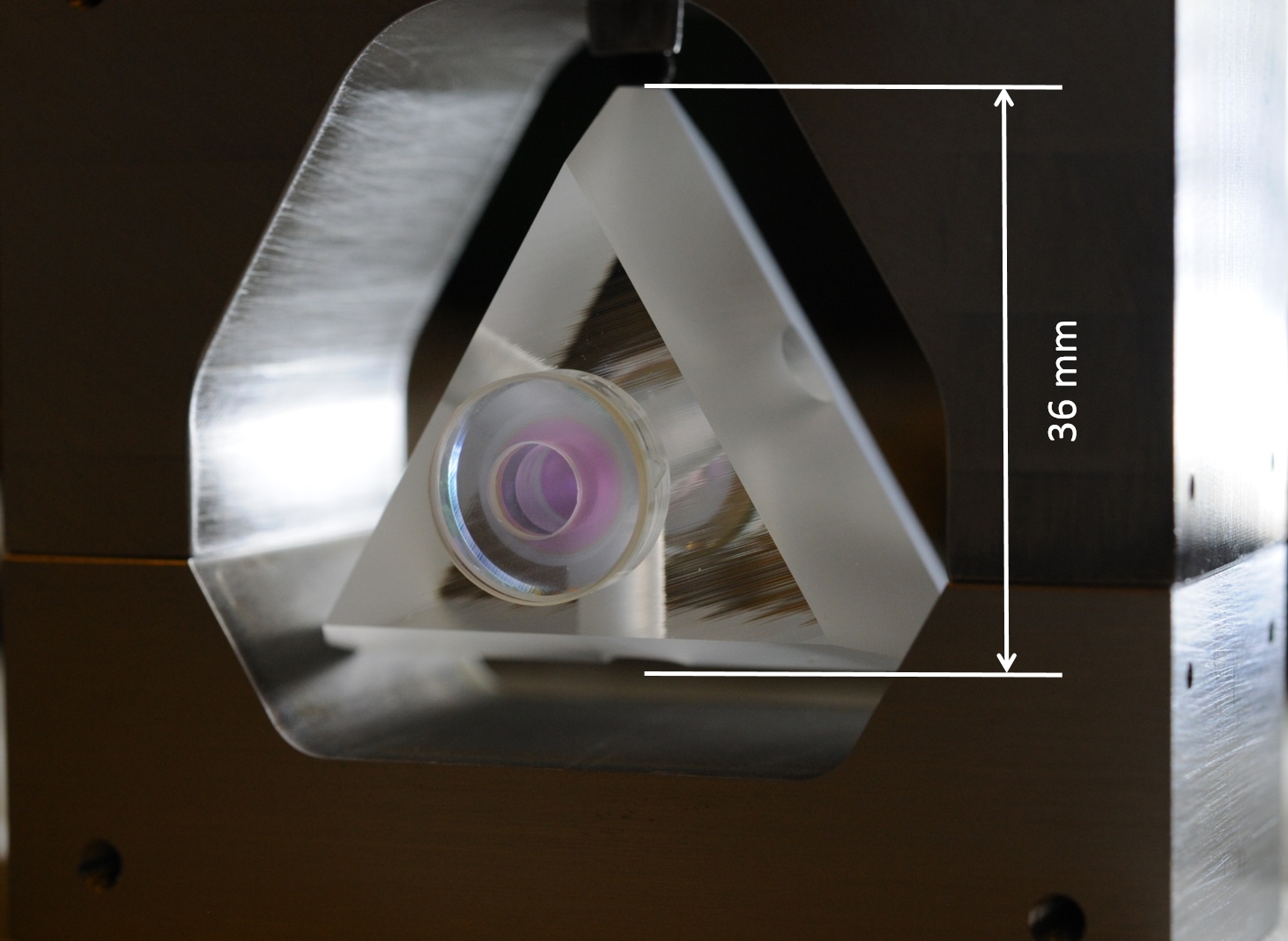}}}
\caption{Example of short FP cavity designs 
\cite{Webster2011, Leibrandt2011a, Davila-Rodriguez2017, Didier2016}.} \label{fig:shortcav}
\end{figure}

Compact cavity designs must rely on the design guidelines mentioned above, with the addition of a robust support and the ability to sustain higher accelerations; moreover, shorter cavity lengths (5 cm or shorter, see examples Fig. \ref{fig:shortcav}) are preferred, in order to reduce the overall volume of the vacuum chamber.

\subsubsection{Cubic cavity design}
Webster and Gill from NPL presented a cubic cavity design, which has the lowest measured passive acceleration sensitivity to date \cite{Webster2011}. The cavity is based on a 5 cm wide ULE cubic spacer and fused silica mirrors. Corner cuts provide flat surfaces to maintain the cavity inside the vacuum chamber. The cavity low acceleration sensitivity relies both on the spacer symmetry, and on the symmetry of the support forces. Here, the cube is maintained by four nylon spheres applying compressive forces on four corner cuts forming a tetrahedron (Fig. \ref{fig:shortcav} (a)). The depth of the cuts $d$ was chosen following finite-element analysis, which showed a cancellation of the cavity length change due to the support forces for $d=6.7$~mm.

The resulting passive acceleration sensitivities are 2.45(3), 0.21(4) and 0.01(1) ${\times} 10^{-11}/g$ in the axial and transverse directions, the lowest published values to date. Even though no fractional frequency stability was reported for this setup, one can expect a thermal noise limit close to $7{\times} 10^{-16}$ for a 5 cm-long ULE cavity with fused silica mirror substrates \cite{Leibrandt2011a}.

\subsubsection{Spherical cavity design} \label{sec:sphere}
A 5-cm diameter spherical cavity design was presented by Leibrandt \emph{et al.} (NIST) in 2011 \cite{Leibrandt2011a}, see Fig. \ref{fig:shortcav} (b). The spacer geometry relies on the perfect symmetry of the sphere and a ``magic angle'' mounting: the cavity is held at two points chosen at an angle from the optical axis which cancels the cavity length dependence to the support forces.

The measured acceleration sensitivities are 4(5), 16(3) and 31(1) ${\times} 10^{-11}/g$ in the vertical and two horizontal directions \cite{Leibrandt2011a}. The measured fractional frequency stability is about $1.2 {\times} 10^{-15}$ between 0.4 and 13~s. This setup was tested on-field by putting the whole system in a car trunk, where the acceleration sensitivity was reduced by an order-of-magnitude using a feed-forward electronic control loop \cite{Leibrandt2011}.

To increase the setup robustness, a second mounting system was developed using Torlon$^{\rm TM}$ balls held by flexure springs in an Invar$^{\rm TM}$ cubic mount. Combining this new mount and a feed-forward correction of inertial forces, Leibrandt \emph{et al.} measured the lowest acceleration sensitivity to date, below $10^{-12}/g$ in all directions, with a fractional frequency stability of $2 {\times} 10^{-15}$ between 0.5 and 10~s \cite{Leibrandt2013}.

A commercial version of this cavity exists, and was used at FEMTO-ST to demonstrate ultra-low noise all-optical microwave generation \cite{Didier2015}.

\subsubsection{Shorter cavities} \label{sec:shortcav}
More recently, two shorter cavity designs have been proposed, with a cavity length of 2.5 cm, in order to further reduce the setup total volume \cite{Didier2016, Davila-Rodriguez2017}.
\\

The cavity designed by Davila-Rodriguez \emph{et al.} \cite{Davila-Rodriguez2017}  at NIST is based on a simple cylindrical ULE spacer with fused silica mirrors and ULE rings \cite{Legero2010}, see Fig. \ref{fig:shortcav} (c). The cavity is held by three elastomer balls on each of the cylinder faces, the balls being placed at positions that minimize the cavity length change due to the support forces. An Invar$^{\rm TM}$ holder rigidly maintains the elastomer balls in place.

The measured acceleration sensitivity is below $4.5 {\times} 10^{-10} /g$ in all directions, and the laser phase noise is limited by the thermal noise of the cavity, leading to a fractional frequency stability of $2 {\times} 10^{-15}$ around 0.1~s. The total optical cavity volume is of 61~mL, allowing for a potentially compact complete setup.
\\

The cavity designed at FEMTO-ST also relies on a 2.5 cm long ULE spacer, with a double-tetrahedral geometry \cite{Didier2016}, see Fig. \ref{fig:shortcav} (d). Fused silica mirror substrates with ULE rings are optically contacted to the spacer, which is held in its middle plane by three Viton$^{\rm TM}$ balls inside a stainless steel holder. The cavity geometry is optimized to minimize the resonator sensitivity to accelerations, with simulated acceleration sensitivities below $10^{-11}/g$ in all directions. The mirrors are realized using crystalline coatings in order to further reduce the thermal noise \cite{Cole2013}, which is estimated at $1.5{\times}10^{-15}$. The cavity fits in a 2.5 L vacuum chamber, and the complete setup volume will be on the order of 50 L. A preliminary measurement shows $\sigma_y(\mathrm{1\,s})=7{\times}10^{-15}$ \cite{Didier2017}.

\subsubsection{Larger cylindrical cavities}
Other transportable cavities have been constructed based on more traditional spacer designs, with excellent frequency stability and compact overall volume.

The first demonstrated transportable ultra-stable laser was developed by PTB and the D\"usseldorf university for a transportable Sr lattice clock \cite{Vogt2011}. ULE mirrors are contacted to a horizontal, 10 cm-long cylindrical ULE spacer held semi-rigidly by four Viton$^{\rm TM}$ pads in a stainless steel mount. The total volume of the setup, including electronics, is about $3.3\ \mathrm{m}^3$. Even with its quite high sensitivity to vertical accelerations (about $2.7{\times}10^{-10}/g$), it reaches a fractional frequency stability of $2{\times}10^{-15}$ at 5~s integration time.

A following approximately cylindrical, horizontal 10~cm-long ULE cavity was designed in the frame of the STE-QUEST proposal, using fused-silica mirror substrates and a refined supporting frame \cite{Chen2014}. With a 24~L optical setup volume, the cavity acceleration sensitivity is below $4{\times}10^{-10}/g$ in all directions, and its fractional frequency stability reaches $1{\times}10^{-15}$ between $1 - 10$~s integration times.

A 10 cm long, vertical transportable cavity was developed at NPL \cite{Parker2014}. The ULE spacer is a tapered cylinder held close to its middle plane using a central supporting collar. Its is bonded to an aluminum support ring using a vulcanizing, ultra-high vacuum (UHV)-compatible silicone sealant. The cavity being used only for optical frequency transfer experiments, it was not temperature-stabilized, and its fractional frequency stability was limited to $6{\times}10^{-15}$ at 1~s integration time by a high-order frequency drift.

A European consortium recently reported the development of an ultra-stable clock laser system for space applications within the SOC2 project \cite{Swierad2016}. The setup is based on a 10~cm-long, vertical cylindrical ULE cavity with fused silica mirrors. The overall volume of the optical setup is below 50 L, with a fractional frequency stability floor at $8{\times}10^{-16}$, close to the calculated thermal noise. The acceleration sensitivity is below $6{\times}10^{-10}/g$ in all directions. This setup follows a previous similar design \cite{Argence2012}.

\begin{table}[b!]
\tbl{Summary of the published passive acceleration sensitivities, fractional frequency stability and volume of portable ultra-stable lasers. All cavities are designed for lasers at 1064 nm, except the ones from NPL and FEMTO-ST at 1.5 $\mu$m and SOC2 at 698 nm. N.p. : not published. Excl. elec.: excluding electronics. Volumes are only indicative, as the published dimensions do not always include the same elements.}
{\begin{tabular}{lccccc} \toprule
 Reference 																													& Cavity length & Max. acceleration sensitivity $/g$	& $\sigma_y$ 						& volume (L)\\ \midrule
 Webster \emph{et al.}, NPL \cite{Webster2011} 											& 5 cm					& $2.45{\times}10^{-11}$ 							& n.p. 									& n.p. \\
 Leibrandt \emph{et al.}, NIST \cite{Leibrandt2011a, Leibrandt2011} & 5 cm					& $3.1{\times}10^{-10}$ 							& $1.2{\times}10^{-15}$ & 81 (excl. elec.) \\
 Davila-Rodriguez \emph{et al.}, NIST \cite{Davila-Rodriguez2017} 	& 2.5 cm 				& $4.5{\times}10^{-10}$ 							&  $2{\times}10^{-15}$  & 0.061 (cavity alone) \\
 Didier \emph{et al.}, FEMTO-ST \cite{Didier2016}									 	& 2.5 cm 				& n.p.									 							&  $7{\times}10^{-15}$  & 50 (excl. elec.) \\
 Chen \emph{et al.}, PTB \cite{Chen2014} 														& 10 cm 				& $4{\times}10^{-10}$ 								& $1{\times}10^{-15}$ 	& 24 (excl. elec.) \\
 Parker \emph{et al.}, NPL 	 \cite{Parker2014}											& 10 cm 				& $2{\times}10^{-10}$ 								&  $6{\times}10^{-15}$  & 300 (excl. elec.) \\
 Swierad \emph{et al.}, SOC2  \cite{Swierad2016}  										& 10 cm 				& $6{\times}10^{-10}$ 								& $8{\times}10^{-16}$ 	& 50 (excl. elec.) \\ \bottomrule
\end{tabular}}
\label{table:cavities}
\end{table}

\section{Ion trap}
All single-ion atomic clocks rely on a variation on the Paul trap \cite{Paul1990}, which uses a combination of DC and RF electric fields to stably trap the ion. As a potential minimum cannot be generated at a point in space with static electric fields alone, the idea of the Paul trap is to combine static and radio-frequency electric fields with amplitudes and frequency such that, depending on the ion mass, the average electric potential seen by the ion is harmonic. There are several ways to implement such a trap, and we will focus only on the geometries relevant to this article.

\subsection{Trap geometries}

\subsubsection{Linear, ring and endcap traps}
The original Paul trap is based on a central ring-shaped RF electrode with hyperbolic cross-section combined with two close-by hyperboloid DC endcaps \cite{Paul1990}. This geometry approximates a harmonic trap to a high order, but offers poor optical access. Three three-dimensional variations on the Paul trap are used in single-ion optical clocks:

\begin{itemize}
	\item \emph{the ring trap}, where conical endcap electrodes are placed further away from the central ring electrode, allowing better optical access. It is used for instance by NIST and PTB for their  Hg$^+$ and Yb$^+$ clocks \cite{Oskay2006, Tamm2000}.
	\item \emph{the endcap trap}, where the ring is removed and two collinear RF endcaps are shielded by external DC electrodes \cite{Schrama1993}. It is used for instance by NRC and NPL for their  Sr$^+$ \cite{Barwood2014, Dube2014}  and Yb$^+$ ion clocks \cite{Godun2014}.
	\item \emph{the linear Paul trap}, derived from Paul quadrupole mass spectrometer. Four parallel electrodes provide the RF trapping potentials while DC endcaps close the trap in the axial direction. This geometry can be used to trap ion chains, and is therefore used in quantum-logic ion clocks \cite{Schmidt2005}, \emph{e.g.} for NIST Al$^+$ clock and PTB Al$^+$ and In$^+$ clocks \cite{Chou2010, Pyka2013}.
\end{itemize}

\subsubsection{Surface electrode traps}
Surface electrode (SE) traps have been proposed and demonstrated in the early 2000s \cite{Chiaverini2005, Seidelin2006}. To generate a 3D trap, they rely on a planar electrode geometry, which can be implemented using simple microfabrication techniques. Their design and production is less complex than 3D structures, and SE traps using simple rectangular electrodes are widely used in the field of Quantum Information Processing (QIP) with trapped ions. Such traps can be modeled analytically \cite{House2008, Wesenberg2008} and enable multiple trapping sites, fast ion transport, single-ion addressing, etc. Their two main drawbacks come from the relatively low RF voltages they tolerate (limited by the electrodes breakdown voltages around 200~V), which limit the trap depth, 
and the proximity of the ion to the surface that can induce higher heating rates, reducing the trap lifetimes with laser-cooling on to about a day.
%
This so-called ``anomalous heating" \cite{Turchette2000} can however be mitigated. First, a $d^{-4}$ dependence to the ion-electrode distance $d$ has been observed \cite{Brownnutt2015}, and whereas QIP ion traps have to operate close to the surface ($d<100$~$\upmu$m) in order to allow fast operations, a single-ion atomic clock based on a SE trap can operate with $d>500\ \upmu$m. Secondly, surface treatments by ion bombardment have been shown to greatly reduce the heating rate \cite{Hite2012, McKay2014}. Finally, this anomalous heating is also lowered at cryogenic temperatures \cite{Labaziewicz2008}, even though this solution is unsuitable for compact clocks.

\subsection{Miniature ion traps} \label{sec:microtraps}

Machined traps such as linear Paul traps or ring electrode traps can be fabricated to mm or sub-mm dimensions. Among the projects listed in the next section, four rely on such traditional geometries with reduced dimensions. These geometries offer the advantage of being analytically modeled and offering traps with low anharmonicity. However, when reaching sub-millimeter dimensions, machining uncertainties and surface qualities start to become comparable to or worst than what can be achieved using micro-fabrication techniques.

\begin{figure}
\centering
\includegraphics[width=0.85\columnwidth]{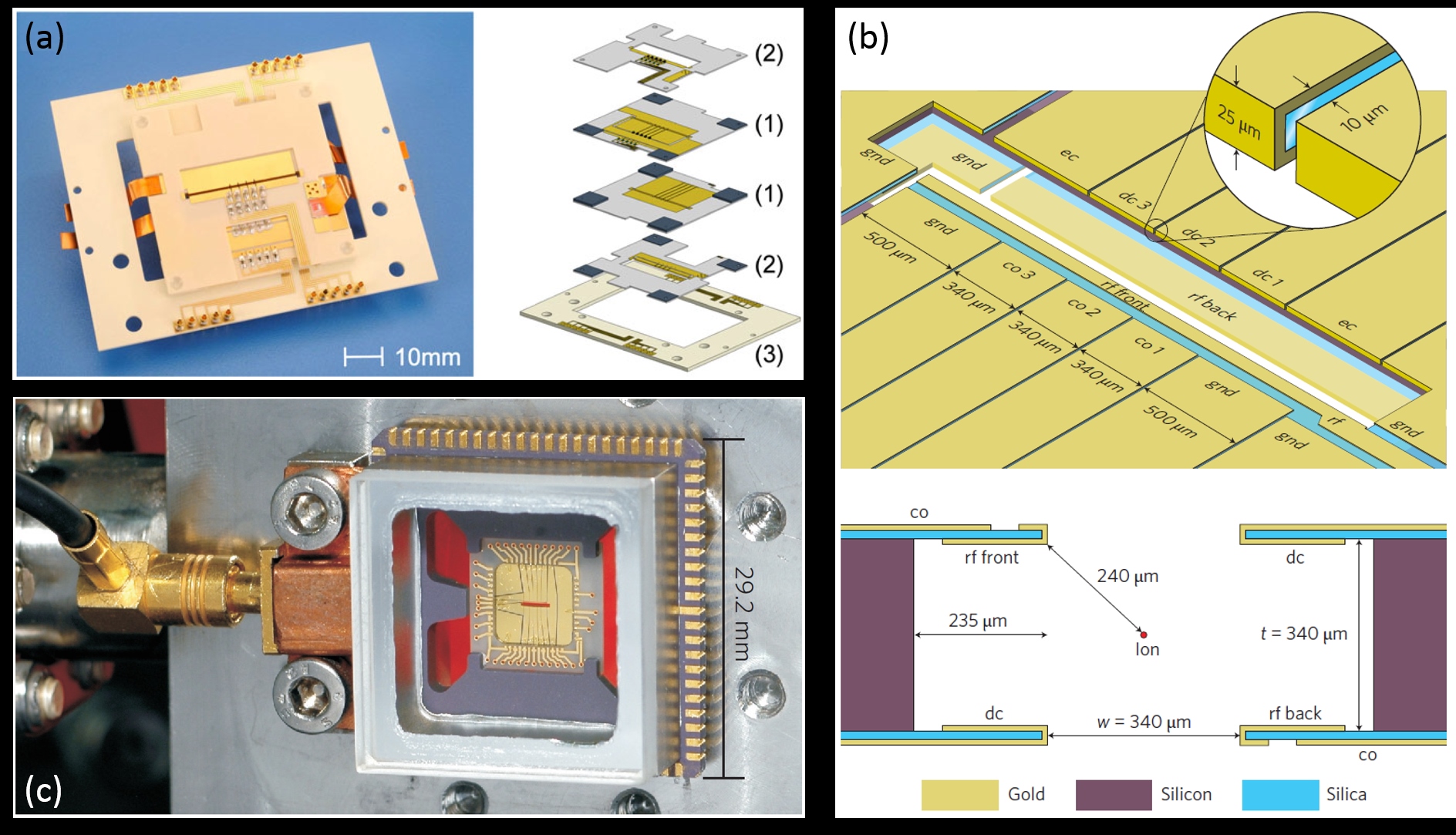}
\caption{Examples of micro-fabricated linear Paul traps. (a) Multi-layer, segmented linear Paul trap for the multi-ion In$^+$ optical clock at PTB, from \cite{Pyka2013}. A similar trap will be used in PTB future Al$^+$ transportable clock. (b) Monolithic, segmented linear Paul trap fabricated on a silica-on-silicon wafer at NPL, from \cite{Wilpers2012}. The design is a multi-ion Sr$^+$ trap, but could well be used in a single-ion atomic clock. (c) Photograph of the NPL trap in its vacuum chamber, with visible air-side electrical access, from \cite{Wilpers2012}.} \label{fig:minitraps}
\end{figure}

Several teams have started developing ion traps in cleanroom facilities, using MEMS fabrication techniques. These technologies combine the advantage of extremely low fabrication tolerances, good surface quality, and potential mass production. Moreover, micro-fabrication processes can be iterated in order to integrate additional functionalities to the trap, including optical elements (see section \ref{sec:outlook}). Such techniques are ideal for producing SE traps, which have rather simple geometries. However, 3D structures can also be realized, and microfabricated linear Paul traps have been demonstrated \cite{Brownnutt2006}.
Fig. \ref{fig:minitraps} illustrates two such 3D micro-fabricated systems, developed at NPL (see description below) and PTB. Many other similar traps have been developed in the field of quantum information computing \cite{Hughes2011}.
\\

A micro-fabricated Sr$^+$ trap with a compact vacuum package was developed at NPL \cite{Wilpers2012, Wilpers2013}. The segmented linear Paul trap is etched from a silica-on-silicon wafer using microfabrication techniques. Gold electrodes are deposited on the insulating silica, while the low-resistivity silicon wafer minimizes RF losses. 
The ion-to-electrode distance of 240 $\upmu$m leads to a heating rate below 320 phonons/s and a 30 min trap lifetime (with cooling off).

The resulting physics package is inscribed inside a 200 cm$^3$ volume including the trap, electronic package, and square viewports. The achieved pressure at the trap location is estimated through background-gas collisions measurements, and is below $10^{-11}$~mbars.
This trap was not specifically designed nor tested for frequency metrology, yet it gathers many of the building blocks required for a transportable optical clock operation, including ultra-high vacuum, large optical access and out-of-vacuum electrical control of the trap. In the next section, we will describe the existing transportable single-ion optical clock projects and the associated miniature traps.


\section{Existing projects}
\label{sec:projects}

To our knowledge, there are at least five existing transportable single-ion optical clock projects so far \cite{MIKES,PTB_QUEST,Brewer2014, Cao2017, Lacroute2016}, with only one with published clock performances. Table \ref{table:ionclocks} summarizes the trap geometries, target or measured volumes and performances, as well as clock laser FP cavity geometry. Another academic project, based on trapped Ca$^+$ ions addressed using integrated optics and a fully-fibered optical setup, has started at the University of Sussex \cite{Sussex}. An additional, industrial project led by Toptica Photonics AG and the PTB was very recently announced, aiming at the realization of a transportable single-ion clock prototype\footnote{www.opticlock.de}.

\begin{table}[b!]
\tbl{Summary of the published existing transportable single-ion clock projects. Syst. unc.: fractional systematic uncertainty. Stab. : fractional frequency stability. Excl. elec.: excluding electronics. N.p. : not published.}
{\begin{tabular}{lcccccc} \toprule
 Laboratory 									& Ion 		& Trap geometry 							& Cavity geometry 					& Overall volume 						& Performances\\ \hline\midrule
 PTB \cite{PTB_QUEST} 				& Al$^+$ 	& Segmented linear Paul trap 	& n.p.											& n.p. 											& n.p.  \\ \midrule
 NIST \cite{Brewer2014} 			& Al$^+$ 	& Linear Paul trap 						& 5 cm spherical spacer 		& A few m$^3$					 			& $8{\times} 10^{-18}$ \\ 
															&   			&     												&     											& (target)									& (syst. unc., target)\\ \midrule
 VTT MIKES \cite{MIKES} 	& Sr$^+$ 	& Endcap trap 								& n.p.											& n.p.   										& n.p. \\ \midrule
 WIPM \cite{Cao2017} 					& Ca$^+$ 	& Ring trap 									& 10 cm 										& 0.54 m$^3$ 								& $2.3 {\times} 10^{-14} \tau^{-1/2}$ (stab.),\\
															&   			&   													&   												& (excl. elec.)							& $7.8{\times}10^{-17}$ (syst. unc.)\\ \midrule
 FEMTO-ST \cite{Lacroute2016} & Yb$^+$ 	& SE trap 										& 2.5 cm tetrahedral spacer & 500 L											& $10^{-15} \tau^{-1/2}$\\
															&   			&   													&   												& (target)									& (stab., target)\\ \hline
\end{tabular}}
\label{table:ionclocks}
\end{table}

\subsection{PTB and NIST Al$^+$ clocks}

\subsubsection{Quantum logic spectroscopy}
Some ions, including Al$^+$ and In$^+$, have interesting metrological properties but do not offer easily addressable transitions for laser cooling and clock state detection. For such species, a ``quantum logic" spectroscopy protocol was proposed and demonstrated in 2005 \cite{Schmidt2005}. It was successfully used by a number of groups since then, and extended to the domain of molecular spectroscopy \cite{Rosenband2008, Leibfried2012, Wolf2016}.

The underlying idea is to trap a spectroscopy ion together with a logic ion which can be used both to sympathetically cool and to read-out the spectroscopy ion energy state. This is made possible by the coherent coupling of the trapped ions motional states through the Coulomb interaction. The trapped ions motional state can be coherently manipulated using Raman pulses, and internal states can be reversibly transferred to motional states  \cite{Meekhof1996}. In the quantum logic spectroscopy protocol, a state of motion is therefore used as a transfer state to coherently map the spectroscopy ion internal state, which cannot be read-out, to the logic ion internal state, which can be read-out \cite{Schmidt2005}.

\subsubsection{PTB Al$^+$}
\label{ssec:ptb}
The PTB Al$^+$ transportable clock project relies on a segmented linear Paul trap, and a clock laser based on a transportable cavity developed at PTB \cite{PTB_QUEST}. The target fractional frequency accuracy is of order $10^{-18}$.  Al$^+$ is interrogated and cooled using the quantum logic spectroscopy method \cite{Schmidt2005}, with Ca$^+$ ions being used as the logic ion. The targeted applications are both on-field, direct clock-to-clock comparisons and chronometric leveling. The performances of the clock should stem from the Al$^+$ ion low sensitivity to light shifts and BBR as well as from the trap design.

The Al$^+$ trap design is based on a geometry developed for PTB multi-ion indium clock \cite{Herschbach2012, Pyka2013} implemented by stacking four segmented electrode layers, see Fig. \ref{fig:minitraps}(a). The trap geometry has been optimized in order to minimize micro-motion along the trap axis, while trapping the ions sufficiently far ($\approx 700 \ \upmu$m) from the electrodes to minimize heating. A study conducted by the team showed that both motional and BBR shifts contributions to the fractional frequency uncertainty could be reduced to below $10^{-19}$~\cite{Keller2016}. To do so, the authors use either photon-correlation or resolved sideband methods to measure and minimize the fractional $2^{\rm nd}$-order Doppler shift caused by micromotion. The secular motion contribution is estimated by cooling the ion close to the motional ground state, and by measuring the heating rate in the absence of cooling. The extremely low heating rate achieved (below 1.8 phonon/s in all directions) translates into a $2^{\rm nd}$-order Doppler shift increase below $3.5{\times}10^{-20}\mathrm{s}^{-1}$ for $^{115}$In$^+$. Finally, the BBR contribution is estimated using both temperature sensors and a calibrated infrared camera to determine the trap temperature distribution. The results are fed to an finite elements method (FEM) simulation software, which determines the temperature at the ion location with a 0.1~K uncertainty, corresponding to a BBR shift fractional uncertainty of $2{\times}10^{-20}$~\cite{Dolezal2015}.

\subsubsection{NIST Al$^+$}
\label{ssec:nist}
The transportable Al$^+$ clock developed at NIST \cite{Brewer2014} relies on a linear Paul trap \cite{Chen2017}. Gold RF electrodes are sputtered on a laser-machined 300~$\upmu$m thick diamond wafer, while macroscopic titanium endcaps close the trap. The ion-electrode spacing is 250~$\upmu$m. The diamond wafer allows a homogeneous temperature distribution over the trap, reducing the uncertainty of the electrodes temperature, and thus reducing the BBR contribution to systematic uncertainties. Moreover, this trap has a remarkably low heating rate, enabling sideband cooling of the ion to its motional ground state, hence reducing the uncertainty on time-dilation shifts to $10^{-19}$. The clock laser is based on the ultra-stable, spherical FP cavity described in section \ref{sec:sphere}. The clock uses quantum logic spectroscopy as well, this time with Mg$^+$ logic ions. Target performances, volume and applications are similar to the PTB design.

\subsection{WIPM Ca$^+$}
The compact Ca$^+$ clock developed by Cao \emph{et al.} at WIPM in China is based on a ring trap of 0.75~mm radius and ion-to-endcap distance \cite{Cao2017}. The trap RF frequency balances the second-order Doppler shift and the RF AC Stark shift, while the BBR shift due to the trap temperature is evaluated at the mHz level using combined measurements and FEM simulations.

\begin{figure}%
\centering
\includegraphics[width=0.48\columnwidth]{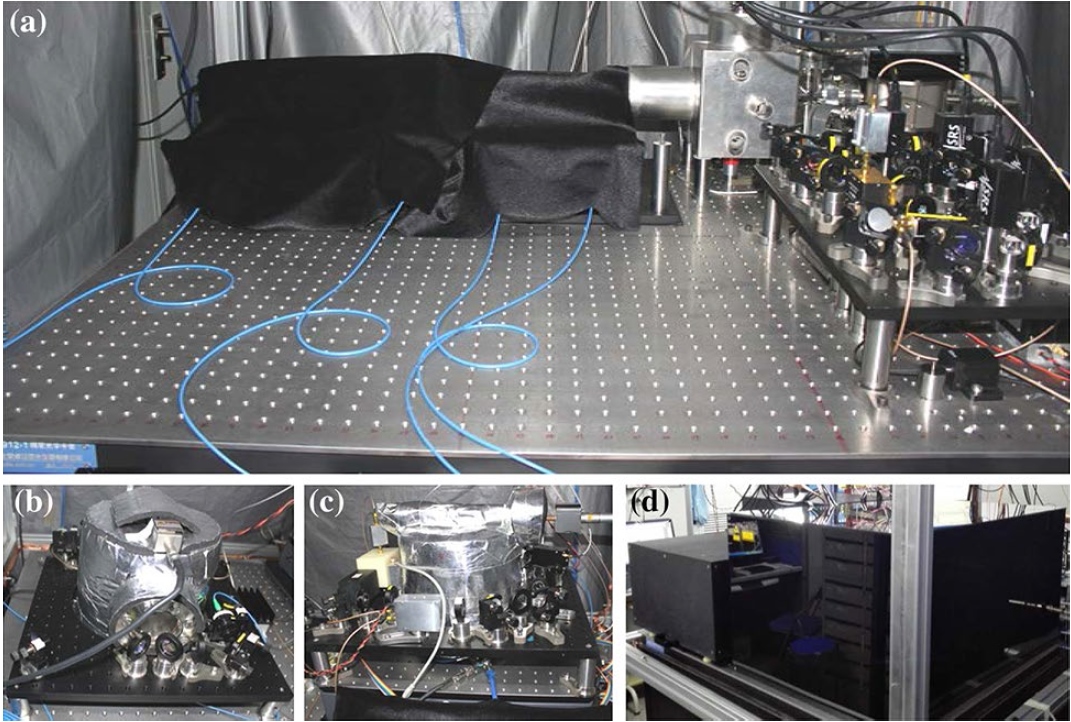}%
\includegraphics[width=0.42\columnwidth]{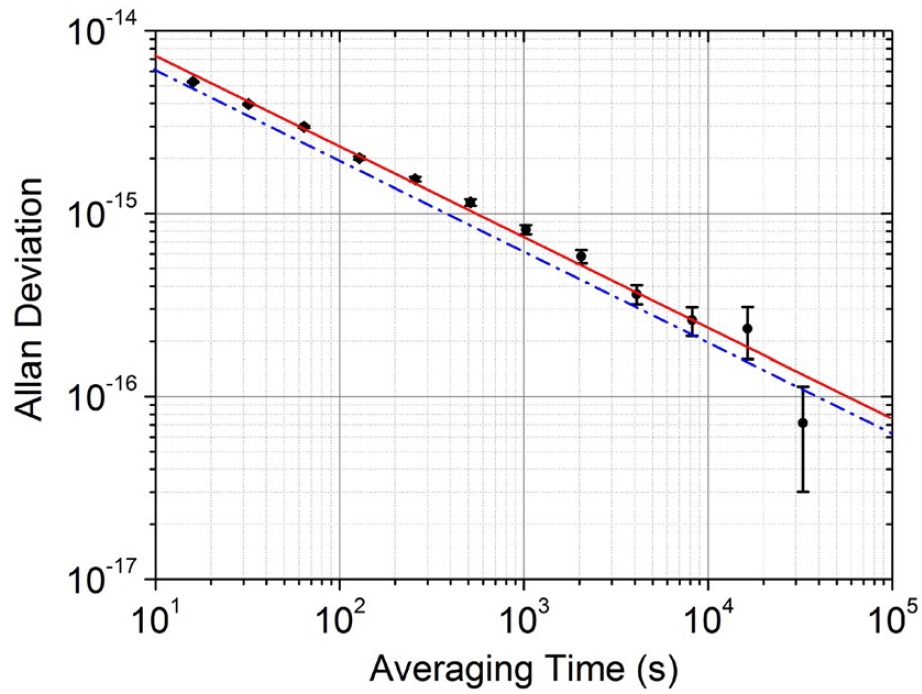}%
\caption{Transportable Ca$^+$ clock at WIPM (from \cite{Cao2017}). \emph{Left:} pictures of the ion trap, laser sources and detection sub-module (a), laser-stabilization cavity (b), ultra-stable clock laser cavity (c), and of the assembled physical system (d) with overall volume 0.54 m$^3$. \emph{Right:} Fractional frequency stability, measured using self-comparison. The measured stability (black data points and fitted red line) is close to the quantum projection noise limit (dashed blue line).}%
\label{fig:cao}%
\end{figure}

The WIPM Ca$^+$ transportable clock project is the most advanced, with a demonstrated fractional systematic uncertainty below $10^{-16}$ in a 0.54 m$^3$ volume (excluding electronics), see Fig. \ref{fig:cao}~\cite{Cao2017}. The ion trap is enclosed in a compact vacuum chamber (343~cm$^3$). The cooling, repumping and quenching lasers are locked to a so-called ``3-in-1" Fabry-P\'erot cavity, while the clock laser is locked using PDH to an ultra-stable FP cavity. All sources are diode lasers, except for a 370~nm LED for photo-ionization. More importantly, the current performance of the clock could be improved by one order of magnitude by improving the clock laser stability as well as by a better determination of the electric quadrupole shift \cite{Cao2017, Huang2016}.

\subsection{VTT MIKES Sr$^+$} 
The transportable Sr$^+$ clock developed at VTT MIKES in Finland \cite{Fordell2015}, in collaboration with PTB and NRC, relies on a miniature endcap trap developed by NRC \cite{Dube2013} and a cavity design from PTB \cite{MIKES}. Compact laser sources and frequency stabilization setups have been designed, allowing for a compact optical setup \cite{Fordell2015}.  The end-cap electrodes are 0.25~mm radius molybdenum wires, with an endcap separation of 0.54~mm.

\subsection{FEMTO-ST Yb$^+$}
At FEMTO-ST, we are currently building a single-ion $^{171}$Yb$^+$ optical clock, with a total target volume of 500~L and a fractional frequency stability of $10^{-14} \tau^{-1/2}$. Such performances would be one order of magnitude better than active hydrogen masers in a similar volume. 
The clock laser will be based on our 2.5~cm cavity design (see section \ref{sec:shortcav}). All laser sources are extended cavity laser diodes, stabilized to a wavemeter \cite{Saleh2015}. The clock laser beam at 435.5~nm is generated by second harmonic generation in a PPLN waveguide \cite{Delehaye2017}.

\begin{figure}%
\centering
\includegraphics[width=0.8\columnwidth]{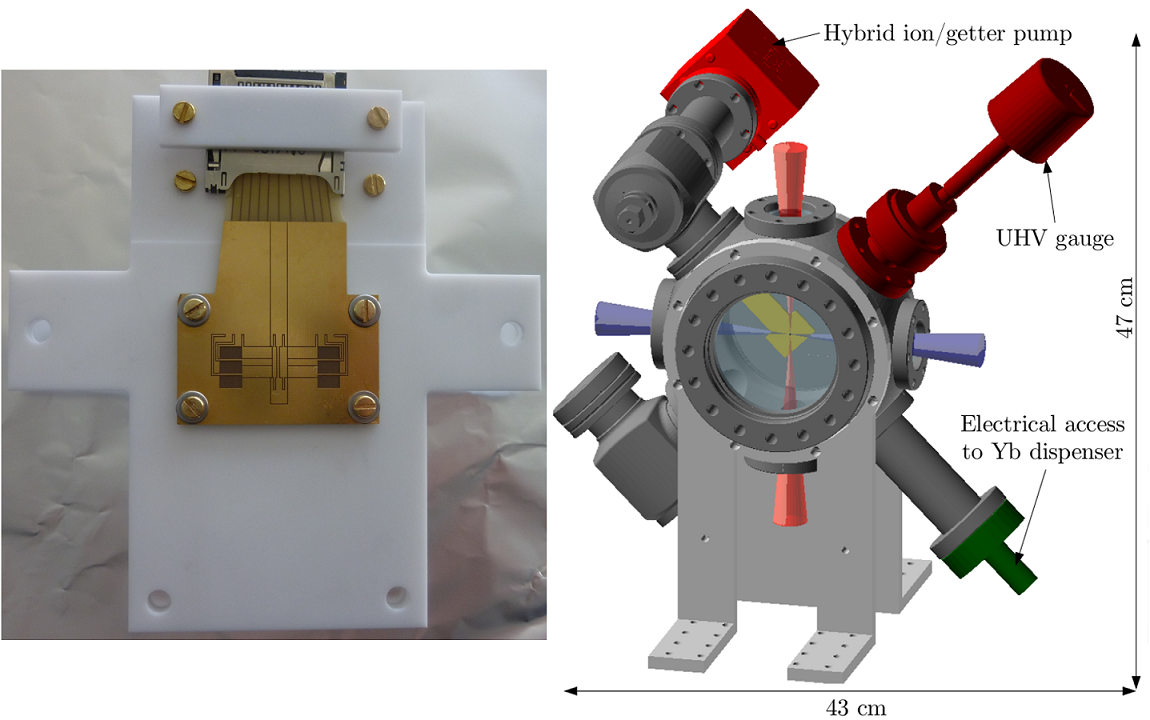}%
\caption{The Yb$^+$ trap setup at FEMTO-ST. \emph{Left:} SE trap mounted on its Macor$^{\rm TM}$ stand. The electrical connections to the chip are based on a mini-SD design that can be seen at the top. See \cite{Lacroute2016} for details. \emph{Right:} octagonal titanium vacuum chamber, with front optical access for detection and peripheral windows for laser cooling and probing. The chamber is pumped using a hybrid getter + ion pump.} %
\label{fig:femtoclock}%
\end{figure}

Our trapping setup is based on a SE trap \cite{Lacroute2016} shown in Fig. \ref{fig:femtoclock}. The trap is designed to confine a single $^{171}$Yb$^+$ about 600~$\upmu$m away from the trap surface. Trapping frequencies superior to 100~kHz in all directions ensure operation in the Lamb-Dicke regime. A first prototype of the trap will be tested in a traditional, compact titanium vacuum chamber illustrated Fig. \ref{fig:femtoclock}. However, SE traps have the potential of a few 100~cm$^3$ vacuum chamber volume, by using the so-called ``atom chips" technique where the trapping chip closes a cubic glass cell pumped to UHV~\cite{Szmuk2015}. The cell provides wide optical access, while electrical connections are kept outside the vacuum chamber. While such setups are often used in the neutral atoms community, it has still to be tested with trapped ions.

\section{Miniaturization perspectives of some other clock components} \label{sec:outlook}

In the previous sections, we have focused on the ``heart" of the clock, namely the ion trap and the clock laser. To further reduce the apparatus footprint, it will be necessary to miniaturize the remaining clock components, which include both control electronics and optical elements. Control electronics can now be made extremely compact thanks to advances in digital electronics. Here, we present some of the emerging technologies which will help miniaturizing optical frequency combs, laser sources and optics for ion trap setups. 

\subsection{Optical frequency combs} \label{sec:comb}

The element that will allow any timing measurements or inter-species comparisons is the optical frequency comb, pioneered by H\"ansch in the early 2000s~\cite{Hansch2006}. The optical frequency comb, also known as femto-second laser, generates a train of ultra-short optical pulses at a rate $f_{rep}$, where $f_{rep}$ lies in the RF or microwave range. When the laser spectrum spans more than an octave, its carrier offset frequency $f_0$ can be measured and stabilized \cite{Telle1999, Jones2000}; when both $f_0$ and $f_{rep}$ are stabilized and referenced, the laser spectrum is indeed a ``comb" with equally-spaced, phase-coherent teeth separated by $f_{rep}$. Each comb component frequency can therefore be determined, and the obtained ``frequency ruler" extends over a wide frequency range. The optical frequency comb can be used, among other applications, to compare different parts of the optical spectrum \cite{Nicolodi2014}, or to link the optical and microwave frequency domains \cite{Fortier2011}. In optical atomic clocks, it is the key element for realizing a so-called ``optical second" \cite{Riehle2015}. For a more detailed explanation of optical frequency combs properties and applications, we point the readers to a few reviews on the subject \cite{Hall2006, Hansch2006, Diddams2010, Hall2017}.

It would therefore be extremely useful for a transportable optical clock to integrate such a comb. Even though these are still voluminous, compact optical frequency combs have already been demonstrated, and miniature combs might be implemented in the near future.

Menlo Systems Inc. have recently demonstrated the operation of a telecom frequency comb inside a sounding rocket~\cite{Lezius2016}. The comb remained mode-locked during the flight, including acceleration and deceleration phases of up to 12~$g$. It was operated in micro-gravity for a total of 360~s, remaining fully phase-locked. The complete setup volume, including electronics, is about 33 L.


Another promising approach, though less mature, is the use of a micro-optical resonator. Toroidal or disc resonators with high quality factors ($Q>10^9$) can exhibit high non-linearities, and octave-spanning frequency comb have been demonstrated using cascaded Kerr effects (see~\cite{Kippenberg2011} for a review).

\subsection{Other lasers}

Apart from the clock laser and the optical frequency comb, at least three additional wavelengths are necessary for all species, and up to five in some cases. Some of these are in the UV range, where sources are scarce. An important post in reducing the clock volume is therefore the laser sources, and several recent advances will allow drastic reduction in the system sizes and power consumptions.

Many laser cooling experiments rely on extended cavity diode lasers (ECDLs) thanks to their narrow linewidths and low power consumption. Recently, miniature ECDLs have been developed by several laboratories and companies, using either micro-resonators~\cite{Liang2015} or integrated optics~\cite{Luvsandamdin2014, Rauch2015}. Moreover, distributed feedback (DFB) or distributed Bragg resonator (DBR) lasers become available at an increasing number of wavelengths; DFBs were for instance developed for a Sr$^+$ clock \cite{Barwood2012}. To reduce the laser sources size and power consumption even further, visible VCSELs were characterized by the Wineland group and frequency-doubled and quadrupled to the UV to perform photo-ionization and laser cooling of Mg$^+$ atoms \cite{Burd2016}.

\subsection{Integrated optics}
Finally, further reduction of the system size could stem from the integration of optical elements to the trap itself. Several demonstrations of such systems have been made in the fields of QIP and cQED (cavity Quantum Electro-Dynamics): among other examples, an optical fiber directly integrated to a surface electrode trap or to an endcap trap allows efficient fluorescence light collection~\cite{VanDevender2010} and characterization of non-classical light fields by the ion~\cite{Takahashi2013}; a transparent SE trap with indium tin oxide (ITO) electrodes allows direct fluorescence light detection by a photodetector placed behind the trap~\cite{Eltony2012}; integration of nanophotonics waveguides and grating couplers below a SE trap allows single-ion addressing from the chip~\cite{Mehta2016a}.

Integration of such tools would be an important step towards miniature single-ion optical clocks.

\section*{Conclusion}

Single-ion optical atomic clocks have benefited from continuous development for the past 40 years \cite{Dehmelt1975}, and now operate with fractional systematic uncertainties on the order of $10^{-17}$ or below. They are mature both in terms of metrological characterization and technical development, and their current performances are relevant to fundamental physics tests, chronometric leveling, the realization of the international atomic time (TAI) and to the current discussions about a redefinition of the SI second. Systematic frequency shifts affecting these standards are now evaluated below $10^{-17}$, including BBR and electric quadrupole shifts. Prospects of fractional systematic uncertainties below $10^{-18}$ using candidates such as Lu$^+$ \cite{Arnold2015}, highly-charged ions \cite{Safronova2014} or nuclear transitions \cite{Peik2003} are unfolding.

Single-ion optical clocks technological building blocks are also mature. Space-qualified ion traps have been demonstrated \cite{Prestage2007}, as well as miniature ion traps with sealed vacuum chambers \cite{Schwindt2016}. Many of the useful transitions wavelengths can be generated using laser diodes; some require second harmonic generation, which can nowadays be performed using commercial, fibered components with high efficiencies.

Transportable atomic clocks using Al$^+$, Ca$^+$, Sr$^+$ and Yb$^+$ are being developed. All combine a single ion trap with a compact ultra-stable laser and a compact optical bench. The Ca$^+$ clock at WIPM has demonstrated a fractional frequency stability of $\approx 2 {\times} 10^{-14} \tau^{-1/2}$ and a fractional systematic uncertainty of $7.8{\times}10^{-17}$ in a 0.54 m$^3$ volume. With the recent development of compact laser sources and compact optical frequency combs, one could even develop a frequency standard with similar performances in a volume under 100 L. Furthermore, efforts directed towards the realization of single-ion space clocks would turn them into new tools to unveil the fundamental laws of the universe.

\section*{Acknowledgements}
The authors would like to thank Yann Kersal\'e for his careful reading of the manuscript, and Jacques Millo and Alexandre Didier who provided the pictures of the spherical and tetrahedric cavities.

\bibliographystyle{tfpmd}
\bibliography{reviewsic}

\end{document}